\documentclass[%
 reprint,
nofootinbib,
amsmath,amssymb,https://www.overleaf.com/project/628d4119e08f4ef301f2c6d2
aps,
pra,
]{revtex4-2}
\usepackage{graphicx}
\usepackage{dcolumn}
\usepackage{bm}
\usepackage{hyperref}
\usepackage[dvipsnames]{xcolor}
\usepackage{booktabs}
\usepackage{array, longtable, booktabs, makecell}
\usepackage{wasysym}
\usepackage{siunitx}

\begin{document}

\preprint{APS/123-QED}

\title{Cryogenics of a superconducting LINAC : SPIRAL2 from commissioning to operation}

\author{Adnan Ghribi$^{1,2}$}
\author{Muhammad Aburas$^{1,3}$}
\author{Pierre-Emmanuel Bernaudin$^{1,3}$}
\author{Patrick Bonnay$^3$}%
\author{François Bonne$^3$}
\author{Frédéric Bouly$^2$}
\author{Marco Di Giacomo$^{1,3}$}
\author{Charly Lassalle$^{1,4}$}
\author{David Longuevergne$^2$}
\author{François Millet$^3$}
\author{Jean-Pierre Thermeau$^2$}
\author{Arnaud Trudel$^{1,3}$}
\author{Quentin Tura$^{1,2}$}
\author{Adrien Vassal$^{1,4}$.}

\affiliation{$^1$Grand accélérateur National d'Ions Lourds (GANIL)}
\affiliation{$^2$Centre National de la Recherche Scientifique (CNRS)}
\affiliation{$^3$Commissariat à l'énergie atomique et aux énergies alternatives (CEA)}
\affiliation{$^4$Université de Caen - Normandie}





\date{\today}

\begin{abstract}
The SPIRAL2 superconducting linear accelerator (LINAC), which has been operational since 2019, employs superconducting, independently phased RF resonators to deliver a wide range of particle beams. Designed for flexibility in particle types, intensities, and energies, it utilizes superconducting quarter-wave resonators (QWRs), whose performances are critically dependent on the reliability of the cryogenic operation. This paper reviews the evolution from commissioning to the routine operation of the SPIRAL2 cryogenic system, initially commissioned in 2017. It highlights the key challenges encountered, including thermo-acoustic oscillations, thermal management, and abnormal behavior of cavities. Furthermore, it explores the integration of thermodynamic modeling and machine learning techniques to enhance system control and diagnose issues. This work serves as a comprehensive resource for advancing the cryogenic operation and performance of superconducting LINACs.
\end{abstract}

\maketitle


\section{\label{sec:1}Introduction}
The SPIRAL2\footnote{https://www.ganil-spiral2.eu} linear accelerator (LINAC) is based on superconducting (SC), independently phased RF (Radio Frequency) resonators. To allow the required broad ranges of particles, intensities, and energies (see Table \ref{tab:beam_spec}), it is composed of two families of short cryomodules developed by the CEA/Irfu\footnote{Commissariat à l'énergie atomique et aux énergies alternatives} (Saclay) and CNRS/IN2P3/IJCLab\footnote{Irène Joliot-Curie Laboratory} teams. The first family, called type A and denoted $A_{i \in [01,12]}$, is composed of 12 quarter-wave resonators (QWR) optimized for $\beta = 0.005$ (one cavity/cryomodule). The second family, called type B and denoted $B_{i \in [01,07]}$, is composed of 14 QWR optimized for $\beta = 0.12$ (two cavities/cryomodules). The resonance frequency was 88.0525 MHz, and the maximum gradient in the operation of the QWRs was $E_{acc} = V_{acc}/\beta\lambda = 6.5$ MV/m. Developed by CNRS/IN2P3/LPSC (Grenoble), RF power couplers provide up to 12 kW of continuous wave (CW) beam loading power to each cavity.

The accelerator has been in operation since August 2019, providing proton and deuteron beams to the Neutrons for Science Facility (NFS). The SPIRAL2 cryogenic system is extensively detailed in \cite{ghribi2017status}. The commissioning of the cryogenic system started in 2017 and has been operating in the nominal operation mode since 2019. This paper aims to provide the community with a general overview of the path from cryogenic commissioning to beam operation. It includes detailed issues such as thermo-acoustic oscillations, thermodynamic modelling, LINAC thermal imaging, machine learning-based virtual observers, dynamic heat load compensation, and abnormal behavior of cavities.

The first section describes the constraints and requirements of cryogenic operation of the SPIRAL2 LINAC. The second section examines the roles of thermodynamic modelling, from control to physics-based and machine-learning-based inferences of cavity thermal dissipation. The third section details the issues faced during commissioning and operation, as well as how they had been diagnosed and mitigated.

\begin{table}
\caption{SPIRAL2 Beam Specifications. Checked boxes are to date achievements.}\label{tab:beam_spec}
\begin{minipage}[b]{.5\textwidth}\footnotesize
\begin{tabular*}{1\textwidth}{@{\extracolsep{\fill} }l*{8}{c}r}
\toprule
Particles  & \multicolumn{2}{c}{$H^{+}$} & \multicolumn{2}{c}{$^{3}He^{2+}$} & \multicolumn{2}{c}{$D^{+}$} & \multicolumn{2}{c}{ions} & \scriptsize{[Units]}\\
\midrule
Q/A	& 1 & \CheckedBox{ }  & 3/2 & \CheckedBox{ } & 1/2 & \CheckedBox{ } & 1/3 & $\Box{ }$ & \\
Maximum current & 5 & \CheckedBox{ } & 5 & \CheckedBox{ } & 5 & \CheckedBox{ } & 1 & $\Box{ }$ & \scriptsize{[mA]}\\
Minimum energy & 0.75 & \CheckedBox{ } & 0.75 & \CheckedBox{ } & 0.75 & \CheckedBox{ } & 0.75 & $\Box{ }$ & \scriptsize{[MeV/A]}\\
Maximum energy & 33 & \CheckedBox{ } & 24 & \CheckedBox{ } & 24 & \CheckedBox{ } & 15 & $\Box{ }$ & \scriptsize{[MeV/A]}\\
Maximum beam power & 165 & $\Box{ }$ & 180 & $\Box{ }$ & 200 & $\Box{ }$ & 45 & $\Box{ }$ & \scriptsize{[kW]}\\
\bottomrule
\end{tabular*}
\end{minipage}
\end{table}

\section{\label{sec:constraints}Constraints and requirements of the cryogenic operation}
Cryogenics allow superconducting cavities to reach the required conditions to inject and maintain radio-frequency (RF) at the required power level, thereby allowing them to operate at their nominal accelerating gradient.

The first requirement is to maintain the superconducting state of the cavities and ensure that their temperature-dependent surface resistance remains uniform. In other words, the system must be able to efficiently extract the RF dissipated heat from the cavity walls. This is achieved by plunging the cavities in a liquid helium bath, also known as a liquid helium phase separator (LHPS). The cavities were completely immersed in liquid helium at all times. If not fully immersed, a cavity may undergo a quench, that is, partial or total loss of its superconducting state.

\begin{table}
\caption{Main causes of perturbations that may affect RF operation of the SPIRAL2 cavities.}\label{tab:perturb}
\begin{minipage}[b]{.5\textwidth}\footnotesize
\begin{tabular*}{1\textwidth}{@{\extracolsep{\fill} }l*{2}{c}r}
\toprule
Source & Frequency & Correction \\
\midrule
Lorentz force & $\sim$ Static & FTS\footnote{Frequency Tuning System} \\
Helium P/L\footnote{Pressure/Level} cross-coupling & $\sim$ Hz &  CS\footnote{Cryogenic System} \\
Helium CM\footnote{Cryomodule} cross-coupling & $\sim$ Hz &  CS \\
Cryogenic operation set-point & $\sim$ Hz & FTS \\
RF coupler temperature & $\sim~10^{-2}$ Hz &  FTS\\
Other vibrations\footnote{Includes thermo-acoustics, vacuum pumps vibrations and helium turbulence.} & $\sim$  Hz $\Rightarrow~\sim$ 100 Hz& LLRF\\
\bottomrule
\end{tabular*}
\end{minipage}
\end{table}

The other operating conditions are linked to any source of perturbations that may disturb the frequency of operation of the cavities. Table \ref{tab:perturb} summarizes most of these sources. Because niobium is not infinitely stiff, any source of vibration in the liquid helium bath induces mechanical deformations and, therefore, frequency shifts. When the oscillations are fast ( $>100$ Hz), the low-level radio frequency (LLRF) system control loop ensures the stability of the accelerating field both in amplitude and phase. However, this ability is limited by the available power margin of the RF amplifiers. The frequency mismatch $\Delta{f}$ with respect to the operation frequency $f_{0}$ of the machine can be written as 

\begin{equation}
\Delta{f} = \frac{f_{0}}{2Q_{L}} \sqrt{ \frac{P_{gmax}}{P_{g0}} - 1}
\end{equation}
where $Q_{L}$ is the loaded quality factor, $P_{g0}$ is the required RF power, $f_{0}$ is the resonant frequency, and $P_{gmax}$ is the maximum power delivered by the RF generator. 

For SPIRAL2, because of the poor amplifier design margins, $P_{gmax}$ is clearly the limiting factor; for example, the threshold is already reached when experiencing helium pressure thermoacoustic oscillations in cryomodule/valve boxes. For the type B cryomodules, the limit in terms of pressure amplitudes was set to 5 mbars for 5 Hz frequency modulations. When pressure oscillations are slow ($< 1$ Hz), it is the role of the Frequency Tuning System (FTS) to compensate frequency instabilities\cite{LONGUEVERGNE20147}. However, there are cases in which the FTS and LLRF correction capabilities fall short. High-amplitude liquid-helium phase-separator pressure instabilities are a good example of phenomena that cannot be corrected by the FTS or compensated by the LLRF. This property is common to all-low-bandwidth and high-quality-factor superconducting accelerating cavities, which are sensitive to micrometric mechanical deformations. This constraint can be expressed in terms of the frequency-pressure sensitivity $S_{p}$, a property that has been measured during the qualification phase of SPIRAL2 cavities \cite{bernaudin2014spiral2}.Table \ref{tab:sensPT} lists these properties. Based on the RF phase shift $\delta \phi$ between the incident power and the transmitted power, it is possible to translate this constraint in term of maximum acceptable pressure variation $\delta P_{bath}$ of the LHPS :

\begin{equation}
\delta P_{bath} = \frac{\delta f}{S_{p}} = \frac{ f_{0}\delta Q }{ \pi Q_{L} S_{p} }
\label{eq.deltapbath}
\end{equation}where $\delta f$ is the admissible frequency shift and $f_{0}$ is the operating frequency or tuning frequency of the cavity. The values of $\delta P_{bath}$, as calculated in equation \ref{eq.deltapbath}, are listed in Table \ref{tab:sensPT} for the measured $S_{p}$ and $\delta \phi = \pi/12$.

\begin{table}[ttb]
\caption{Pressure sensitivities of SPIRAL2 cavities for $\delta \phi=\pi/12$ et $Q_{L}\sim10^{6}$.}\label{tab:sensPT}
\begin{minipage}[b]{.5\textwidth}\footnotesize
\begin{tabular*}{1\textwidth}{@{\extracolsep{\fill} }l*{2}{c}r}
\toprule
Cavities\footnote{Numbered by positions in the direction of the beam} & Cryomodules & $S_{p}$ [Hz/mbar] &  $\delta P_{bain}$ [$\pm~mbar$] \\
\midrule
01 & A01 & -1.23 & 11.92\\ 
02 & A02 & -1.32 & 11.11\\
03 & A03 & -1.453 & 10.09\\
04 & A04 & -1.45 & 10.11 \\
05 & A05 & -2.9 & 5.05\\
06 & A06 & -1.08 & 13.58\\
07 & A07 & -1.66 & 8.83\\
08 & A08 & -1.24 & 11.82\\
09 & A09 & -1.38 & 10.62\\
10 & A10 & -1.31 & 11.19\\
11 & A11 & -1.58 & 9.28\\
12 & A12 & -1.22 & 12.02\\
\midrule
13 & B01 & -5.3 & 2.76\\
14 & B01 & -4.95 & 3\\
15 & B02 & -5.9 & 2.48\\
16 & B02 & -7.3 & 2\\
17 & B03 & -4.9 & 3\\
18 & B03 & -5.2 & 2.82\\
19 & B04 & -5.2 & 2.82\\
20 & B04 & -4.5 & 3.25\\
21 & B05 & -5.9 & 2.48\\
22 & B05 & -6.2 & 2.36\\
23 & B06 & -6.2 & 2.36\\
24 & B06 & -5.8 & 2.52\\
25 & B07 & -5.4 & 2.71\\
26 & B07 & -5.8 & 2.52\\
\bottomrule
\end{tabular*}
\end{minipage}
\end{table}

\section{The roles of thermodynamic modelling}

\subsection{Model description\label{sec:model_desc}}

We have shown how cryogenics can affect RF operation and, in turn, the beam operation. The constraints highlighted in Section \ref{sec:constraints} depend on several localized and distributed systems and sub-systems. One might consider a level 0 subsystem with its physical components as a given constraint that matches a single process variable. For example, cryomodule/valve-box level one sub-system would be made of two level zero subsystems for liquid helium level and helium pressure control, respectively. A level two subsystem would correspond to a given section of the LINAC. An even upper level system would, in that case, include the entire cryoplant with complexity increasing as the number of parameters to be controlled increases. Knowledge of the different subystems can pprovide several benefits:
\begin{itemize}
\item Give a better visibility of the thermodynamic phenomena occurring in the LINAC.
\item Avoid unwanted behaviors thanks to fast reactive and predictive controls.
\end{itemize}

For a level one subsystem simplified model corresponding to a single-cavity cryomodule with its valves-box, the main components to be modeled appear to be two valves for the helium inlet and outlet, the liquid-helium phase separator, and the dynamic heater that mimics the behavior of the RF dynamic heat load. This simple model, documented in \cite{vassal2019dynamic} and based on the Simcryogenics\cite{bonne2020simcryogenics} library, permits a good prediction of the thermodynamic state of the system.The linearisation of the physics model around the operation set-point allows an efficient tuning of the valves controllers (see \ref{sec:control}) and the generation of a real-time heat-load observer.

\subsection{Optimized cryogenic control\label{sec:control}}
One consequence of the described model is that the liquid helium level and phase separator pressure can be predicted to an acceptable level. This, in turn, allows us to tackle one of the main issues that cryogenic operation has been experiencing, namely, helium pressure control. This difficulty is related to the fact that the time constant of the valves is on the order of a second, while helium pressure, for instance, can vary much faster (up to 20 Hz for non-periodic variations and up to 100 Hz for thermo-acoustic oscillations).
At the beginning of the project, the corresponding process variables were managed using two separate and independent Proportional Integral (PI) controllers. The main problem with these controllers is that they are single-input/single-output. Having a working model of this first level 0 subsystem allowed us to synthesize and optimize a different one with multiple inputs/outputs. The new controllers we used are based on the well-known linear quadratic controllers (LQ) \cite{mcgee1985discovery, torres2020kalman}. These controllers were validated on the LINAC through a separate test Programmable Logic Controller (PLC) before deployment in the main cryomodules PLCs.

\begin{figure}
\includegraphics[width=8.5cm]{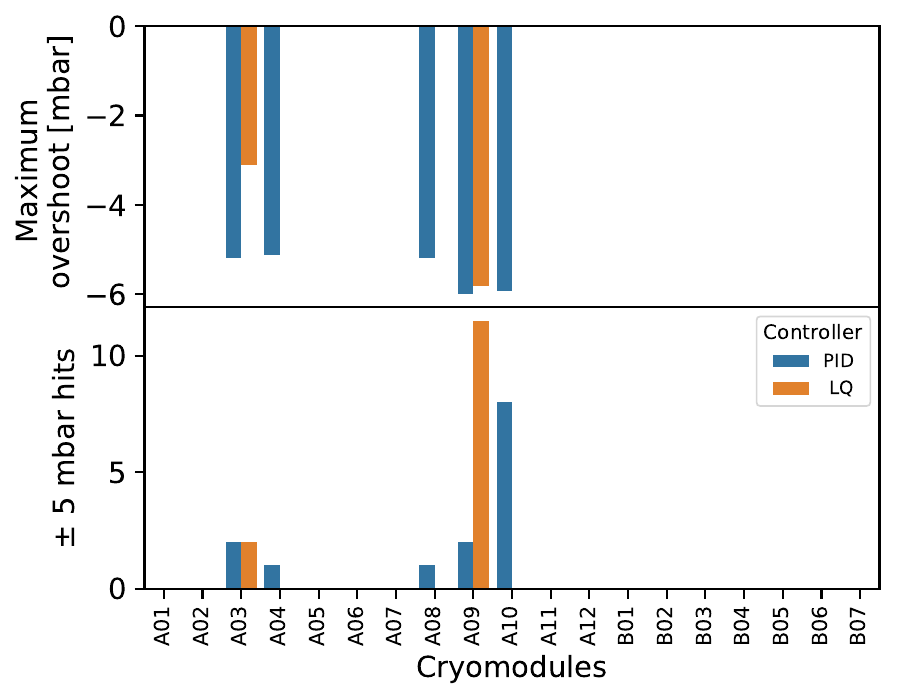}
\caption{PID (optimized) vs LQ (unoptimized) pressure overshoot for deviations $\Delta (P)>\pm5~mbars$.\label{fig:pt_dev}}
\end{figure}

Liquid helium level control being less critical than pressure (see section \ref{sec:constraints}), the main purpose of deploying the new control algorithm was to better control the pressure-induced frequency detuning of the cavities. When set to strictly withstand the $S_p$ limit (see Table \ref{tab:sensPT}), the PI controllers seemed to perform sufficiently well for type A cryomodules. For RF regulation of type A cryomodules, the requirement was set to 5 mbar. Software surveillance to monitor pressure overshots higher than 5 mbar showed the advantage of non-optimized LQ on the LINAC with respect to an optimized PID for most cryomodules (see figure \ref{fig:pt_dev}). After the LQ controller optimization at thermodynamic set points matching the required accelerating fields, the monitors did not show any overshoot.

\subsection{Physics based inference of cavities heat loads}

State observers are valuable tools for monitoring the state of a system. These observers may be based on a combination of direct system measurements. However, knowledge of a model offers the possibility of unlocking hidden features. It also opens the gate to the use of physics-oriented machine-learning techniques for state observer generation. In the case of the thermodynamic model described in \ref{sec:model_desc}, a direct consequence is the ability to monitor the total heat load of the cryomodules/valves-boxes subsystems. This relies on the knowledge of two main quantities: helium density variation $\dot{\rho}$ and internal energy dynamics $\dot{u}$ expressed as follows:

\begin{equation}
\dot{\rho} = \frac{\dot{m}_{in} - \dot{m}_{out}}{V_{sep}}
\label{eq:rho}\end{equation}

and

\begin{equation}
\dot{u} = \frac{ \dot{m}_{in} \cdot (h_{in} - u) - \dot{m}_{out} \cdot (h_{out} - u) + Q_{h} }{\rho \cdot V_{sep}}
\label{eq:u}\end{equation}

where $\dot{m}_{in}$ and $\dot{m}_{out}$ are the input and output helium mass flows, respectively, $V_{sep}$ is the liquid helium phase separator volume, $h_{in}$ and $h_{out}$ are the input and output enthalpy, $u$ is the internal energy, and $Q_{h}$ is the total heat load. Equations \ref{eq:rho} and \ref{eq:u} allow direct recovery of the total dissipated heat load $Q_h$.

\begin{table}[ttb]
\caption{Heat load observers comparison.}\label{tab:obs}
\begin{minipage}[b]{.5\textwidth}\footnotesize
\begin{tabular*}{1\textwidth}{@{\extracolsep{\fill} }l*{1}{c}r}
\toprule
Observer type\footnote{Generated with a Luenberger observer.} & Response time [s] & Noise [W] \\
\midrule
Slow			& 1200	& 0.5 \\
Fast 			& 50		& 12 \\
Fast filtered\footnote{A second order low pass filter has been used here.}	& 100	& 2 \\
\bottomrule
\end{tabular*}
\end{minipage}
\end{table}

Such observers are inherently non-linear, and their implementation in automation and control systems can therefore be challenging. To overcome this challenge, linear state observers such as the Extended Kalman Filter \cite{mcgee1985discovery, torres2020kalman} and Luenberger observer \cite{zeitz1987extended} have been used. However, the resulting states are usually narrow-band around the linearization point, that is, the operating conditions. For this reason, the final observer uses the nonlinear model to calculate the density $\rho$ and internal energy $u$ and the linearized model to calculate the matrix gain of the state observer (see \cite{vassal2019etude} for more details on the gain matrix). Three strategies were tested on the model before its application to the real system: (1) a slow observer, (2) a fast observer, and (3) a fast-filtered observer. From the results in Table \ref{tab:obs}, it is clear that there is a trade-off. A fast-filtered observer was considered for the implementation of the system. In addition, calibrations under a wide range of operating conditions were used to reduce errors. 

\begin{figure}
\includegraphics[width=8.5cm]{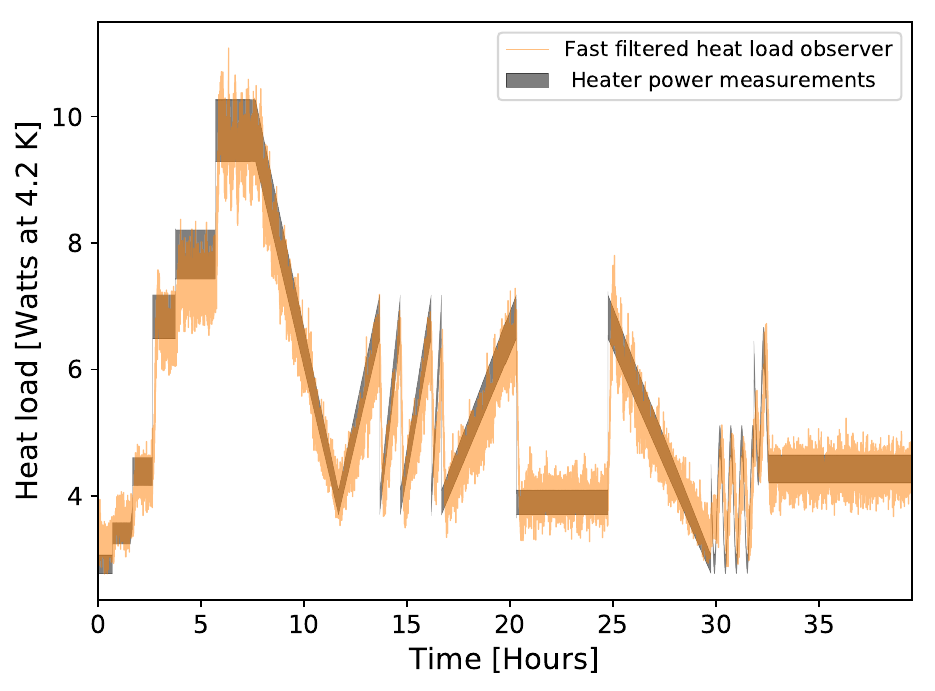}
\caption{Fast filtered heat load observer variation as a heater power is being varied with a specific pattern.\label{fig:hl_observer}}
\end{figure}

The Figure \ref{fig:hl_observer} shows the test implementation of a real-time heat-load observer in a SPIRAL2 cryomodule. The observer is challenged by applying different power variation shapes to a heater that is thermally coupled to a liquid-helium phase separator. The test lasted for more than 40 h, and the heater power amplitudes varied between 0 and \qty{10}{\watt} at \qty{4.2}{\kelvin}. The error related to the applied heater power was considered by calibrating the heater current with the actual heat loads measured by liquid helium decay. The results in the Figure \ref{fig:hl_observer} show an error almost within the precision of the liquid-helium-decay technique. For our application, a heat load measurement by liquid helium decay takes up to \qty{40}{\minute} depending on the power being measured, and cannot be performed during RF or beam operation. The tested observer offers such in situ capability with noise varying between \qty{0.6}{\watt} and \qty{1}{\watt} at \qty{4.2}{\kelvin} and a time response of approximately \qty{10}{\minute}. There is an important margin of improvement in the applied algorithms by tweaking filters and time constants to the monitored behaviors. For example, a \qty{10}{\minute} time response provides good monitoring of the absolute total heat load, while a few seconds time response provides insight into fast abnormal behaviors (relative variations) at the cost of increased noise. One example of abnormal behavior that has been detected is a virtual imbalance between the input and output helium mass flows. In reality, the measured imbalance is a signature of an abnormal gas temperature at the output helium ports. This was caused by the method used to damp cryogenic thermoacoustic oscillations. The identification of the anomaly resulted in the application of a more suitable thermoacoustic damping solution (see Section \ref{sec:thermoacoustics}). Further studies with respect to the beam currents at different duty cycles are required to assess a more complete set of state observers. A complementary approach for which a study is planned will take advantage of beam diagnostics, vacuum gauges, and RF measurements to generate multiphysics observers. Another application under study is the in-situ measurement of the quality factors $Q_{0}$ of the accelerating cavities and the identification of possible degradation causes (field emitters, surface resistance degradation, etc.) \cite{weingarten2011field, LONGUEVERGNE201841}. These applications are coupled with complementary ongoing studies that take advantage of machine learning techniques for advanced diagnostics.

\subsection{Machine learning based inference of cavities heat loads}
\begin{figure}
    \includegraphics[width=8.5cm]{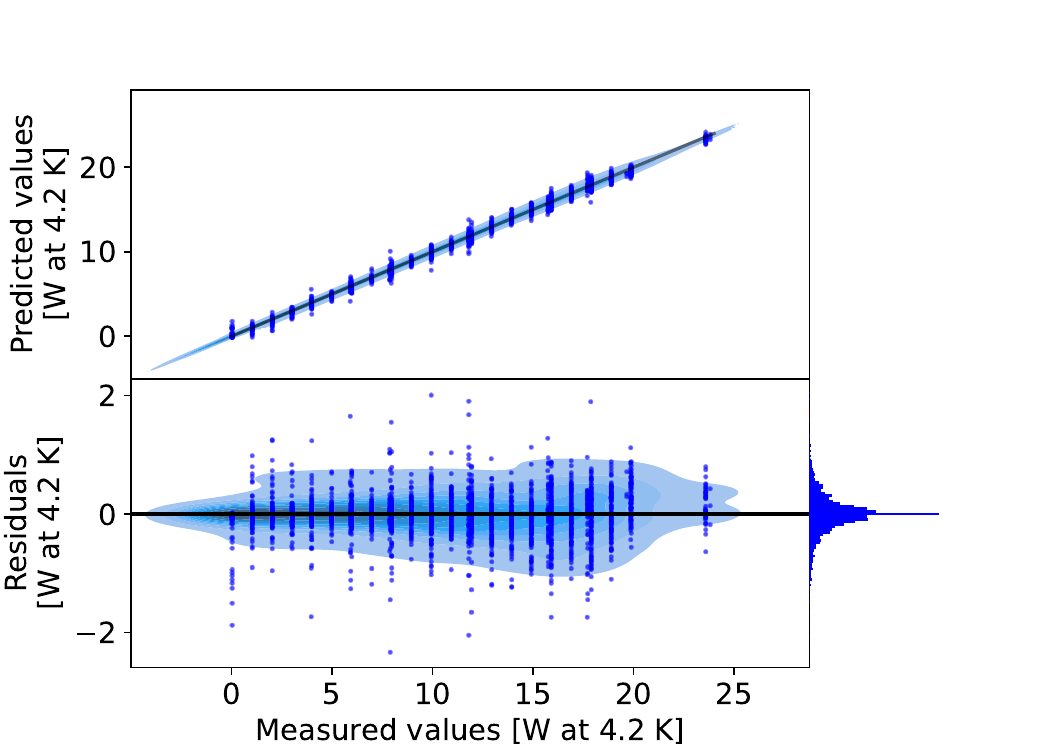}
    \caption{Heater-induced heat load predictions with Stacking model on A08 cryomodule 2022 data at 4.2 \unit{\kelvin}. (top) predicted versus measured values with kernel distribution; (bottom) residuals versus measured values with kernel distribution and distribution of residuals.}
    \label{fig:stackingmodelPLTeTsVeVs_cma082022}
\end{figure}

Machine learning techniques using neural-network-based architectures can offer an interesting alternative to overcome some issues that can arise in model-based observers. However, these methods require a critical quantity of high-quality datasets. To collect data, several machine studies have been performed by inducing a controlled heat load using electric heaters thermally connected to the cavity phase separators. The training phases followed a stepped increase in the induced heat loads with step durations and amplitudes varying from \qty{1}{\minute} to \qty{15}{\minute} and \qty{1}{\watt} to \qty{10}{\watt}. Several neural network architectures have been tested, including Long Short-Term Memory (LSTM) networks with and without attention, convolutional neural networks (CNN), and multilayer fully connected perceptrons (MLP), with several selections of input variables. Another architecture (stacking), created by stacking pre-trained models, was trained using a different training set. For each step, sequences of 10 time steps were created without overlapping and added to the training (40\% for MLP, CNN, LSTM, and LSTMa models + 40\% for the stacking model), validation (10\%), and test (10\%) sets. Table \ref{tab:MLobservers_cma082022evaluationsstatistics} presents the statistics of residuals for these models trained and evaluated on A08 cryomodule 2022 data, with two temperatures, the inlet and outlet valve openings, and the level and pressure of the helium bath\footnote{Temperatures and opening valves are normalized with values at \qty{0}{\watt} and \qty{20}{\watt} induced heat load; for the level and pressure of the helium bath, deviations from the set points are used.}. The models were tested on the same dataset. The Figure \ref{fig:stackingmodelPLTeTsVeVs_cma082022} shows the heat load at \qty{4.2}{\kelvin} predicted using the stacking model and the selection of input variables. The computation time required to predict 2841 values is approximately \qty{0.35}{\second}\footnote{Operating System: Ubuntu 22.04.5 LTS, Memory (RAM): \qty{15}{\giga\byte}, processor (CPU): Intel i5-1235U, Tensorflow version: 2.15.0}.

Further studies with randomly generated heat load steps (amplitude and duration) should be considered. The data can be divided chronologically, and consequently, the length of the sequences can be increased. This would likely lead to improvements in the robustness and accuracy of the models during transients.

\begin{table}
	\caption{Statistics of residuals (in \unit{\watt}) for models trained and evaluated on A08 cryomodule 2022 data at 4.2 \unit{\kelvin}.}\label{tab:MLobservers_cma082022evaluationsstatistics}
    \begin{minipage}{.5\textwidth}
\begin{tabular*}{1\textwidth}{@{\extracolsep{\fill} }l*{5}{c}}
			\toprule
			& MLP & CNN & LSTM & LSTMa & Stacking \\
			\midrule
			Mean & -0.011 & -0.037 & -0.012 & -0.027 & -0.009 \\
			Standard deviation & 0.514 & 0.378 & 0.415 & 0.378 & 0.348 \\
			1st quartile & -0.219 & -0.192 & -0.174 & -0.185 & -0.142 \\
			Median & -0.008 & -0.033 & -0.004 & -0.019 & -0.004 \\
			3rd quartile & 0.211 & 0.133 & 0.172 & 0.135 & 0.133 \\
			Interquartile range & 0.429 & 0.324 & 0.345 & 0.319 & 0.275 \\
			\bottomrule
		\end{tabular*}
        \end{minipage}
\end{table}

\section{From cryogenic to beam operation}

\subsection{Thermo-acoustics : hints, effects and ways out \label{sec:thermoacoustics}}

\begin{figure}[t]
\includegraphics[width=8.5cm]{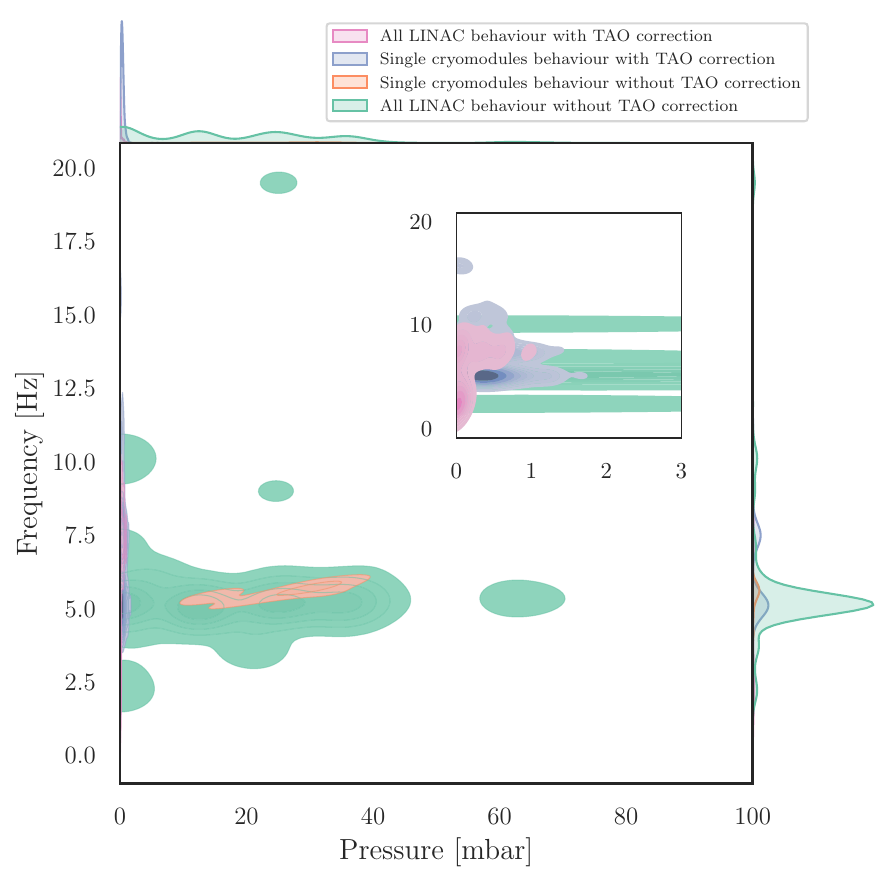}
\caption{Amplitudes and frequencies kernel distributions of detected TAO oscillations for different configurations in the SPIRAL2 LINAC. Zoom window: zoom region for low amplitude oscillations.\cite{ghribi2022cryogenic}.\label{fig:tao}}
\end{figure}

SPIRAL2 has served as a valuable platform for investigating and addressing the challenges posed by cryogenic thermoacoustic oscillations (TAOs) \cite{ghribi2022cryogenic}. These oscillations were observed at multiple locations across a frequency range of 5–100 Hz. Although TAOs in the main phase separator of the cryoplant have been effectively damped using an optimized capacitance, those occurring within the LINAC have proven to be more complex to mitigate. These oscillations, originating in the cryogenic valve boxes and other components of the cryomodules, disrupt the stable LINAC operation by inducing pressure and temperature fluctuations, cavity de-tuning, and increased heat loads. Initial investigations identified the presence of TAOs within the cavities through fluctuations in RF signals and pressure sensor data, excluding external mechanical vibrations as the root cause.

This challenge provided an opportunity to utilize SPIRAL2 as a testbed for studying distributed and interconnected cryogenic thermo-acoustic oscillators, analyzing their impedance properties, and evaluating various damping strategies. Two experimental approaches were used. First, the oscillation amplitudes and frequencies were characterized under nominal operating conditions. Simultaneous piezoelectric measurements of the relative pressure in the liquid helium phase separators of the cavities facilitated precise time- and frequency-domain analyses, resulting in the first thermo-acoustic map of an operational superconducting LINAC (see Figure \ref{fig:tao}). Their analysis indicated that local factors predominantly influenced oscillations when confined to a single cryomodule, whereas more complex cross-coupling effects emerged during simultaneous oscillations across multiple cryomodules.

Second, an adjustable resistance, inductance, and capacitance (RLC) resonator is employed to investigate the acoustic impedance and evaluate the damping strategies. Several damping solutions were subsequently tested, including short-circuit lines, buffers, pistons, and RLC resonators. Among these, the short-circuit lines connecting the 5 K helium gas return circuit with the cavity phase separators demonstrated the highest damping efficiency. However, this approach introduces side effects, such as condensation and valve instability. Adjusting the exchange flow rate between the circuits effectively mitigated these side effects.

One notable advantage of the SPIRAL2 LINAC setup is its ability to actively induce or suppress thermo-acoustic oscillations, thereby functioning as a unique laboratory for studying and mitigating these phenomena in superconducting accelerators.

\subsection{Heat load measurements and automatic RF load compensation}

\begin{figure}[t]
\includegraphics[width=8.5cm]{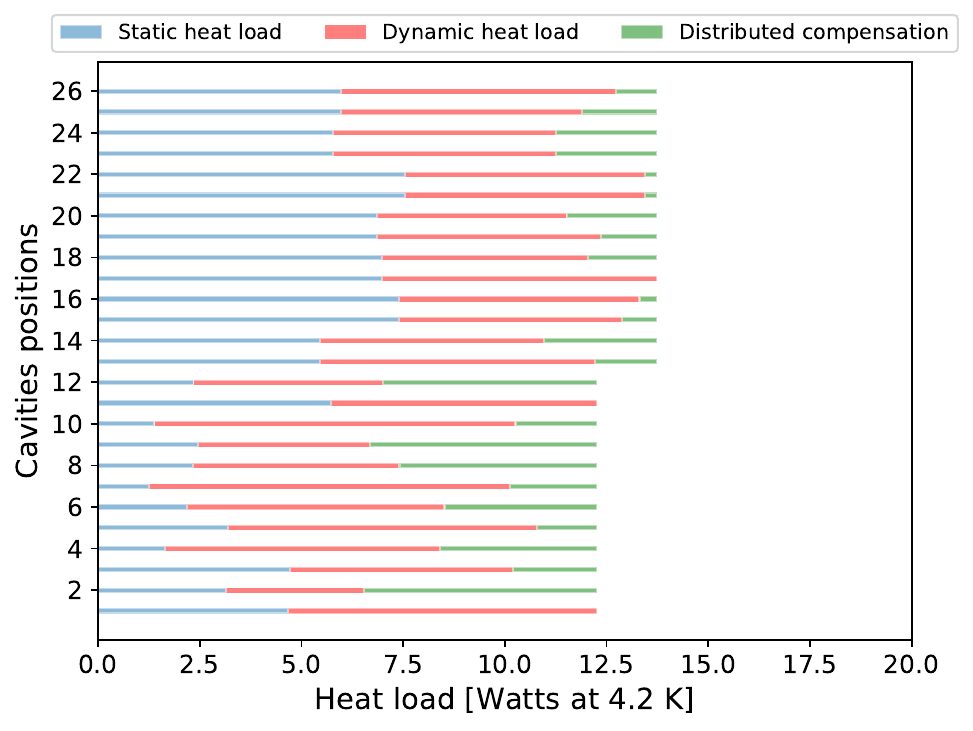}
\caption{Static, dynamic and distributed heat load compensation. The figure shows the actual measurements and compensations applied to the SPIRAL2 LINAC. Abnormal features are hidden for a better visibility and are show in the Figure \ref{fig:heatmap_linac}.\label{fig:hl_comp}}
\end{figure}

Having a thermal image of the LINAC under nominal operating conditions is an important step in reliable cryogenic, RF, and beam operations. In a simplified thermodynamic-electrical analogy, the LINAC can be considered as an unevenly distributed parallel impedance. The cold box acted as the generator. The cryomodules, mounted in parallel, are then operated at the same voltage (or helium pressure), but are subject to different currents (helium mass flows) depending on their characteristic impedances (heat loads). A cryomodule showing a more important heat load can then unbalance the distribution and deprive a neighboring cryomodule of part of the liquid helium flow or cause important instabilities. To avoid this type of problem, the helium flow is equally distributed between the two arms of the LINAC (see the introduction for the shape of the LINAC) and between the different cryomodules. The Figure \ref{fig:hl_comp} shows a snapshot of the heat-load balancing for the SPIRAL2 LINAC. This is done owing to the heaters in the cryomodules, allowing for artificial\footnote{not induced by RF or beam-loading.} heat load compensation. 

For each cryomodule, the total heat load $P_{total}$ can then be expressed as 
\begin{equation}
P_{total} = P_{d} + P_{s} + P_{comp}\label{eq:Pt}
\end{equation}
where $P_{d}$ denotes the dynamic heat load, $P_{s}$ denotes the static heat load, and $P_{comp}$ denotes the compensation heat load for a given cryomodule.

If a reference heat load $P_{ref}$ is considered as the maximum cryomodule heat load in the LINAC without compensation, the compensation heat load for the other cryomodules can be expressed as :
\begin{equation}
P_{comp} = P_{ref} . c - P_{s} - P_{d} \label{eq:Pcomp}
\end{equation}
where $c = 1$  for type B cryomodules and $c = 7/12$ for type A cryomodules (see the introduction for types A and B cryomodules).

\begin{figure}
\includegraphics[width=8.5cm]{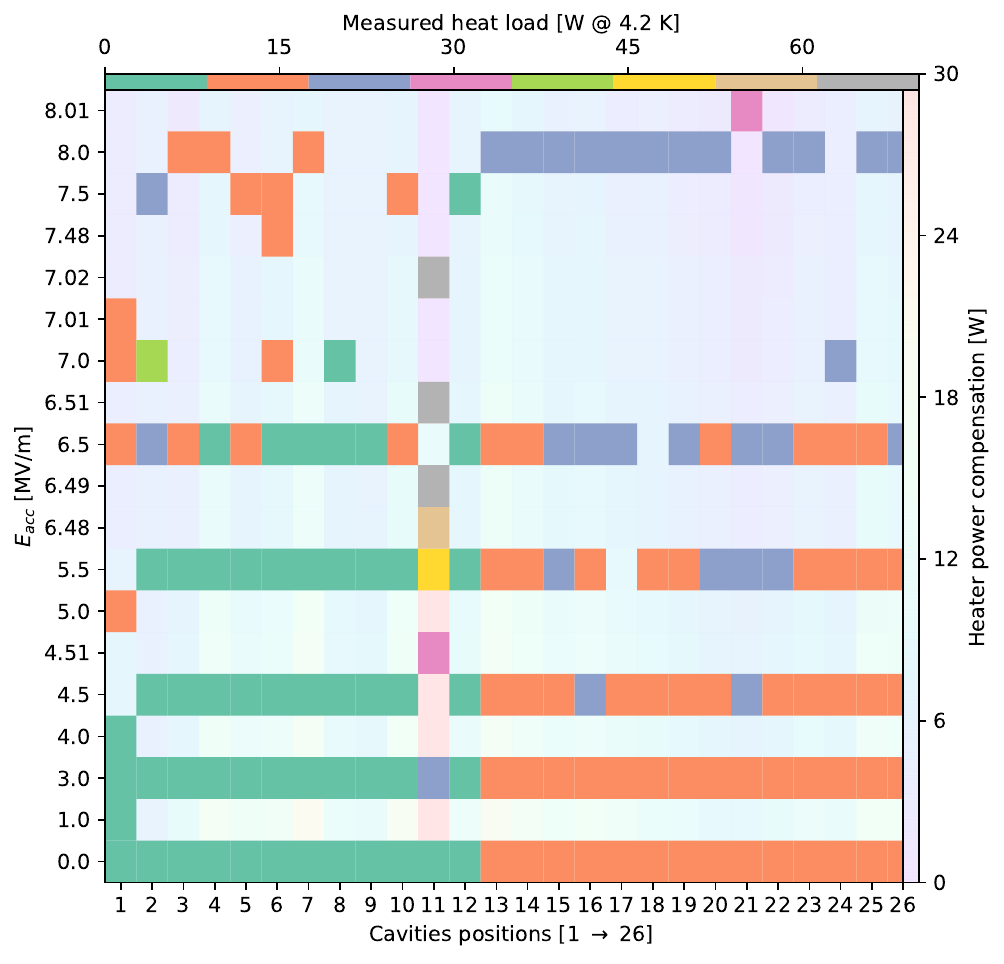}
\caption{Heatmap of the measured LINAC thermal image at different accelerating fields (darker colours) together with the fitted continuous compensation power (lighter colours).\label{fig:heatmap_linac}}
\end{figure}

One consequence of this compensation strategy is that the heat load is kept constant at all times. As the dynamic heat load is variable and depends mainly on the accelerating field of the cavities, knowledge of the thermal image of the LINAC as a function of the accelerating field is an important input. To gain this knowledge, systematic measurements of thermal dissipation were performed at different accelerating fields for every cavity. For this purpose, the liquid helium variation method was used: the inlet valve was closed and the outlet valve was kept under pressure regulation. The liquid helium level decay was then measured. The heat load is given by 
\begin{equation}
P_{thermal} = \frac{dV}{dt}\rho C_{l}
\end{equation}
where $V$ is the liquid helium volume, $t$ is the time, $\rho$ is the liquid helium density at liquid helium pressure, and $C_{l}$ is the latent heat of liquid helium. In practice, the phase separator volume, as a function of the liquid helium probe measurement, was extracted from the 3D models of the cryomodules. $\rho$ was directly calculated from the pressure process value using the CoolProp Python library \cite{doi:10.1021/ie4033999}. Finally, $\frac{dV}{dt}\rho$ was fitted. The entire measurement procedure was automated and integrated into command control. The resulting thermal image of the LINAC is shown in the Figure \ref{fig:heatmap_linac} top axis scale heatmap.

Without dynamic compensation within the cryomodules or helium mass flow balance between the different cryomodules, the dynamic thermal dissipation $P_{d}$ for one cavity can be expressed as follows:
\begin{equation}
	P_{d}= P_{r} + P_{em} = \frac{E_{acc}^{2}}{Q_{0}K^{2}} \label{eq:Pd}
\end{equation}
where $P_{em}$ denotes the extra losses due to field emission (X-rays production caused by electron acceleration in RF field and deceleration in cavity walls). $P_{r}$ represents the effects of the cavity wall surface resistance, $Q_{0}$ is the unloaded quality factor, and K a geometrical constant.

For SPIRAL2, the accelerating cavities operate at accelerating fields lower than 7 MV/m, where field emission effects can be neglected. At these accelerating fields, the so called quality factor drop due to field emission
can be also neglected. This leads to a simple approximated second-order polynomial dependence of the heat load on the acceleration field.
\begin{equation}
	P_{d} = AE_{acc}^{2} \label{eq:Pto}
\end{equation}
with $A=1/Q_{0}K^{2}$.

\begin{figure}
\includegraphics[width=8.5cm]{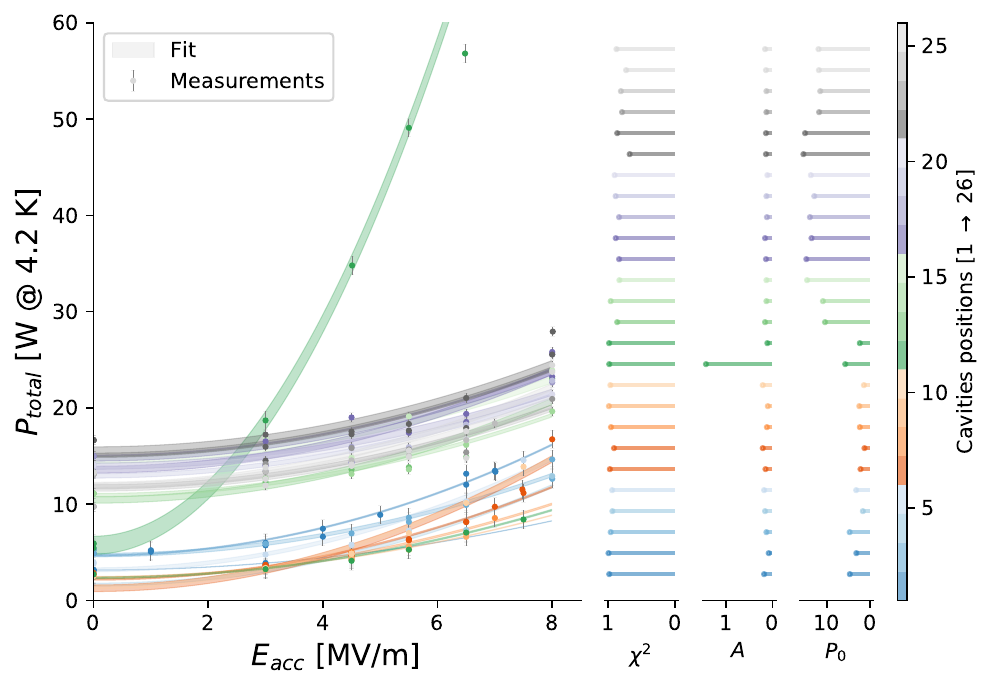}
\caption{Measured and fitted heat load dissipation of the SPIRAL2 cavities as a function of the accelerating gradients. $P_{0}$ represents the heat load dissipation at zero accelerating gradient and $\chi^2$ the goodness-of-fit.\label{fig:PEacc}}
\end{figure}

The Figure \ref{fig:PEacc} shows the measurements of the total heat loads of the cavities as a function of the accelerating fields fitted to Equation \ref{eq:Pto}. Apart from the cavity 11 anomaly (discussed in Section \ref{sec:cma11}), one can clearly differentiate two groups corresponding to the two cryomodule families. The main reason for this is that the two groups exhibit different static heat loads.

Consequently, the heater compensation power needed to keep the heat load constant and balanced when the accelerating field varies can be expressed as:
\begin{equation}
P_{comp} = A(E^{2}_{acc,nom} - E^{2}_{acc,meas}) + P_{dist}
\end{equation}
where, $P_{dist} = P_{ref}.c - P_{s}$ ( Equation \ref{eq:Pcomp}). The calculated power compensation heatmap for $E_{acc,nom} = 6.5$ MV/m is shown in Fig. \ref{fig:heatmap_linac}.

\subsection{In situ cavities anomalies diagnostic}

When anomalies arise, equation \ref{eq:Pto} is no longer valid, and neglected terms must be fully accounted for. In equation \ref{eq:Pd}, $Q_{0}$ is no longer constant \cite{weingarten2011field} and is expressed as 
\begin{equation}
	Q_{0} = \frac{G}{R_{s}} \label{eq:Q0}
\end{equation}
where $G$ is a geometric factor, and $R_{s}$ is the surface resistance, expressed by 
\begin{equation}
R_{s} = R_{res} + R_{BCS} \label{eq:Rs}
\end{equation}
where $R_{res}$ is the residual surface resistance, and $R_{BCS}$ is the BCS\footnote{Bardeen-Cooper-Schrieffer resistance} resistance. $R_{res}$ is, in turn, expressed as 
\begin{equation}
R_{res} = R_{0} + E_{acc}R_{acc} \label{eq:Rres}
\end{equation}
where $R_{0}$ is the residual resistance at the null field and $R_{acc}$ is the sensitivity of the resistance to the accelerating field $E_{acc}$.

\begin{figure}
\includegraphics[width=8.5cm]{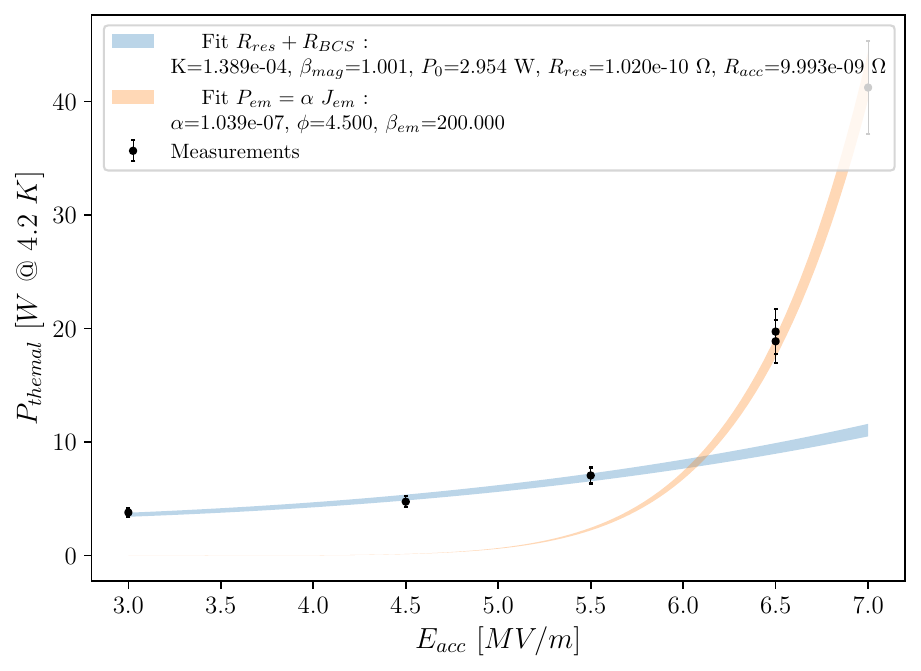}
\caption{A02 cavity heat load dissipation as a function of the accelerating gradient with fits of the surface resistance and the field emission contributions. $\phi$ is a fitted constant, $\alpha$ is proportional to the surface emission and $\beta_{em}$ is a coefficient that multiplies the electric field.\label{fig:Ptcma02}}
\end{figure}

In equation \ref{eq:Rs}, $R_{BCS}$ is a function of the energy gap, temperature, frequency, maximum magnetic field Regarding electronic field emission, $P_{em}$ can be expressed as 
\begin{equation}
P_{em} = \alpha J_{em} \label{eq:Pem}
\end{equation}
with $\alpha$ a constant proportional to the surface emission and $J_{em}$ the Fowler-Nordheim current density \cite{fowler1928proc}.

From equations \ref{eq:Pt}, \ref{eq:Pd}, \ref{eq:Rres} and \ref{eq:Pem}, it follows for a given cavity, temperature and frequency :
\begin{equation}
\begin{array}{ccc}
	P_{total} & = & \Xi_{1}(R_0, R_{acc}) E_{acc}^{2}\\ 
	~ & + & \Xi_{2}(R_0, R_{acc}) E_{acc}^{3} \\
	~ & + & P_{0} + \alpha J_{em}
\end{array}
\end{equation}
where $\Xi_{1}$ and $\Xi_{2}$ are functions that depend only on $R_{0}$, $R_{acc}$ and $\beta_{mag}$. An important consequence is that heat load monitoring across a wide range of accelerating fields can allow the diagnosis and differentiation of different symptoms of abnormal cavity behaviors. Measurement campaigns allow the monitoring of parameters such as $\alpha$, the characteristic of field emitter size, or $R_{res}$ for the surface resistance. One example is shown for A02 cryomodule in the Figure \ref{fig:Ptcma02} where the blue curve represents regular behavior typical of SPIRAL2 cavities. In the same figure, above 6 MV/m, the heat load exhibits a shape typical to field emission. A consequence of this abnormal behaviour was to limit the operation of this cavity to low accelerating gradients. This choice might not be possible for high-beta cavities or cavities at the junction of high- and low-beta cavities. This case is discussed in section \ref{sec:cma11}.

\subsubsection{The special case of cryomodule A11\label{sec:cma11}}

\begin{figure}
\includegraphics[width=8.5cm]{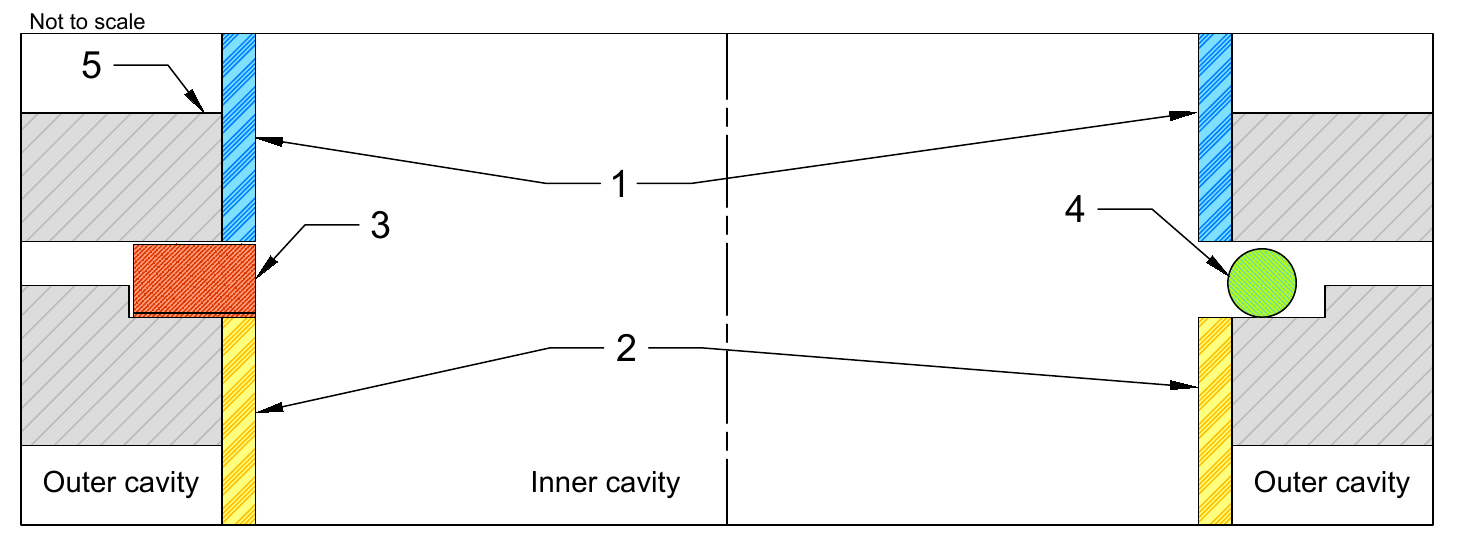}
\caption{cavity bottom sealing techniques used for A11 cryomodule (right part) and all other type A cryomodules (left side). Legend : (1) Inner cavity walls - bulk niobium part ; (2) Inner cavity walls - copper part ; (3) Indium-Copper seal (other type A cryomodules) ; (4) Helicoflex™ seal ; (5) stainless steel flange.\label{fig:joint_cma11}}
\end{figure}

The A11 cavity was the first to be assembled and qualified at the CEA/IRFU, but also the last to use a Helicoflex™ vacuum seal for the removable bottom of the cavity (see Figure \ref{fig:joint_cma11}). These seals generated more heat load than expected and were replaced in all other type A cryomodules by a copper ring and two indium wires acting both as vacuum and RF seals. In addition, the RF power coupler used was prepared and conditioned using non-optimized procedures \cite{martinez2017final}, which were then implemented for all other cryomodules of the SPIRAL2 accelerator. The first qualifications of the cavities on its test stand showed an abnormal behavior typical to a field emitting cavities with heat loads up to 40 W at 4.2 K and 6.5 MV/m. Helium processing \cite{knobloch1998explosive} has been applied and has been efficient enough to cure this cavity and bring its heat load down to 20 W at 4.2 K and 6.5 MV/m, well within the specifications.\vspace{0.1cm}\\

\begin{figure}
\includegraphics[width=8.5cm]{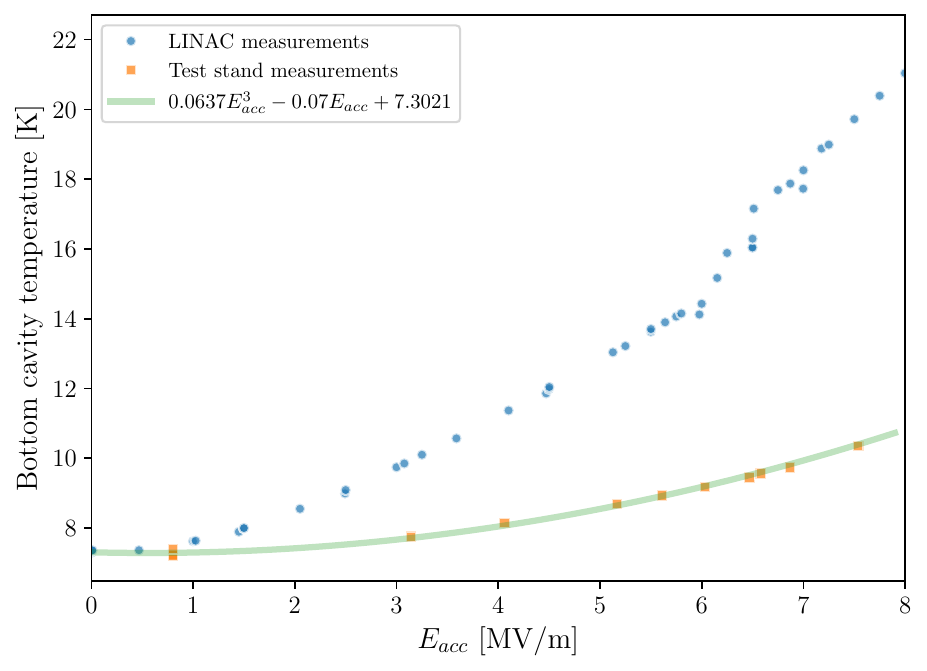}
\caption{A11 cavity's bottom temperature measurement in the LINAC and in the test stand.\label{fig:cma11_tt201}}
\end{figure}

\paragraph{In situ investigations}~\vspace{0.1cm}\\~
However, the first operation in the LINAC showed completely different behavior. The cavity's bottom temperature showed a different dependence on the accelerating gradients, with a jump around 6 MV/m (see Figure \ref{fig:cma11_tt201}). In 2020, a dedicated machine study was set up to localize abnormal heat loads. Cryomodules of the A family are equipped with two helium level sensors, which independently monitor the helium level inside the cavity stem and between the helium vessel and outer cavity walls (see \cite{bernaudin2013assembling} for a detailed description of cavity geometries). The main conclusion of the machine study is that the thermal losses are located either at the bottom of the cavity, on the RF coupler, or on the outside walls of the coaxial cavity. Interestingly, abnormal dissipation did not occur in the main RF loss areas (i.e., the top of the stem and the top torus of the cavity). Combined measurement of heat loads from 1 MV/m to 8 MV/m (see Figure \ref{fig:cma11_htl}) showed three main regions of field dependence losses :\vspace{0.1cm}

\begin{figure}
\includegraphics[width=8.5cm]{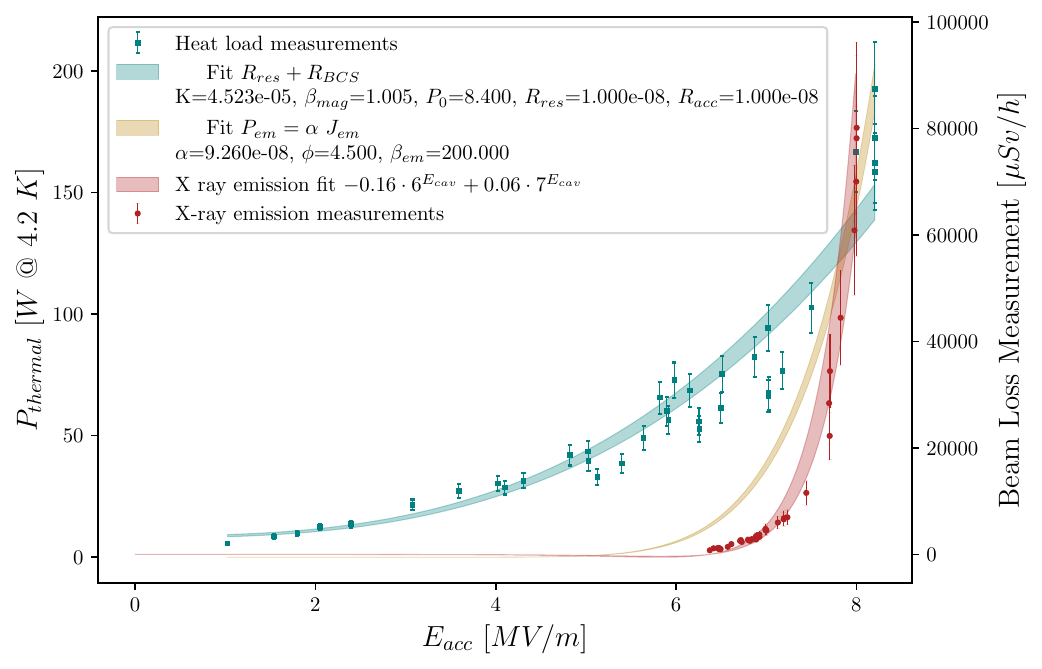}
\caption{A11 cavity heat load dissipation measurements (green points) as a function of the acceleration gradient (left axis) with both the surface resistance (green) and the field emission (yellow) contributions fitted separately. The right axis measurement shows x-ray emission measurements (red) as a function of the accelerating gradient.\label{fig:cma11_htl}}
\end{figure}

\noindent \textbf{1.} The first region is below 5.5 MV/m where the thermal losses seem to be dominated by surface resistance losses. The green curve fit in the Figure \ref{fig:cma11_htl} shows surface resistance $R_{res}$ and resistance field sensitivity $R_{acc}$ up to two orders of magnitude higher than those of the other cryomodules. Two main conclusions were drawn from this observation. The first is that the high sensitivity to the field points toward conductive behavior typical of normally conductive materials. Second, there is a significant degradation in the surface resistance. The latter is an effective value that takes into account three contributions: the surface resistance of the superconducting bulk niobium (typically $\sim n\Omega$), the surface resistance of copper (bottom part of the cavity), and finally, an interface region between the two parts of the cavity centered around the sealing (see Figure \ref{fig:joint_cma11}), but possibly expanding and transiting part of the Nb region to its normal state. However, these contributions are weighted by the magnetic field geometric distribution, which is close to its minimum in the sealing regions. A discontinuity in the conductive inner wall region of the cavity can cause RF field leaks that can drastically increase the RF losses at the interface region.\vspace{0.1cm}

\noindent \textbf{2.} A second region between 5.5 MV/m and 6.5 MV/m where we can see a jump in the temperature of the bottom of the cavity and where the measured heat load seems to flatten.\vspace{0.1cm}

\noindent \textbf{3.} The third region is above 7 MV/m, where the thermal load is dominated by field emission. In this region, the heat load is correlated with the X-ray emissions measured using dedicated beam loss monitors (see Figure \ref{fig:cma11_htl}). This behavior is not problematic because the cavity is not operated above 6.5 MV/m.\vspace{0.1cm}

The three regions described above may correspond to transitions in the liquid helium flow regime due to an increased heat load. The first region likely represents single-phase liquid flow. The second corresponds to the Onset of Nucleate Boiling (ONB), where small vapor bubbles begin to form on heated surfaces. Finally, the third region is dominated by the Leidenfrost effect \cite{walker2010boiling}. In this regime, also known as the Film Boiling Transition (FBT), a very high heat flux leads to the formation of an insulating vapor layer between the liquid helium bath and the cavity surface, significantly reducing heat transfer efficiency and exacerbating the temperature rise.\\ \vspace{0.1cm}

\begin{figure}
\includegraphics[width=8.5cm]{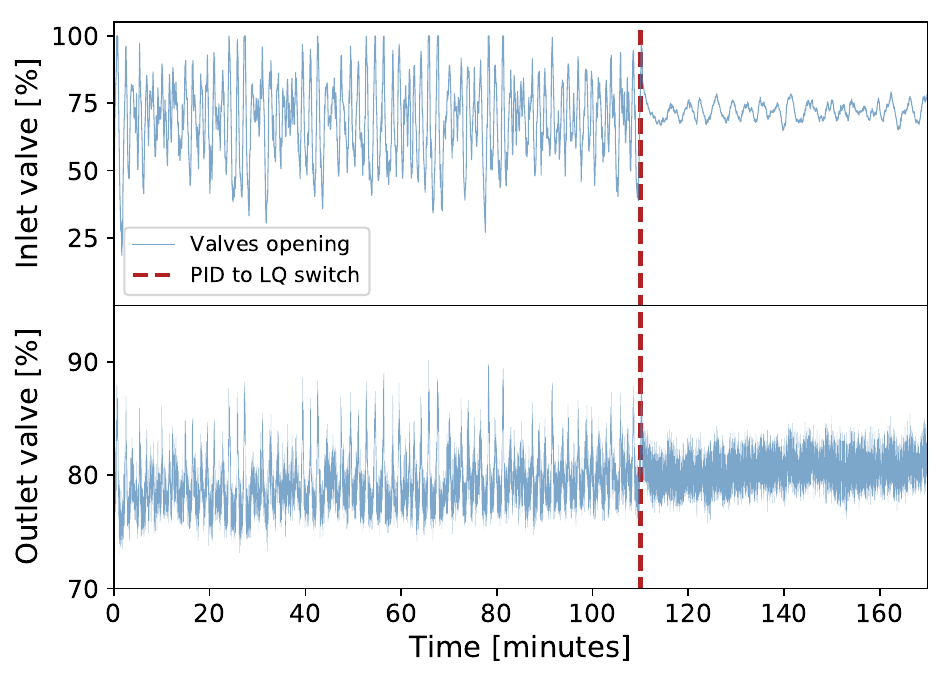}
\caption{A11 cryomodule helium inlet and outlet valves openings before (PID control) and after applying the model based (LQ) control. These measurements were done before changing the CV of the outlet valve.\label{fig:cma11_valve}}
\end{figure}

\paragraph{Operation constraints and solutions}~\vspace{0.1cm}\\~
Given the obtained results, the only way to lower the abnormal heat load of the cryomodule is to disassemble it completely and re-process the cavity in a clean room, using a power coupler prepared using the standard, optimized procedures, and replacing the Helicoflex™ seal with the standard copper + indium assembly. Considering the lack of spare cryomodules and the unavailability of a dedicated test bench for cryomodules in GANIL, this operation was postponed to a later date. The issue with a cryomodule dissipating so much is that the output helium flow is close to the acceptable limits of piping geometries. Consequently, the outlet and inlet valve openings were higher than 75\% when the cavity was operated at 6.5 MV/m. Apart from having valves that operate close to their non-linear range (which adds instability to their control), such openings increase the sensitivity of the helium bath to its surrounding environment (neighboring cryomodules and cryodistribution) and the cross-coupling between liquid helium regulation and pressure regulation. The model-based control detailed in Section \ref{sec:control} allows the effective overcoming of these issues by stabilizing the valves around an optimized operation set point (see  \ref{fig:cma11_valve}). The applied set-point\footnote{The set-point is an optimized heat load around which the model is linearized before extracting the valves control matrices} was chosen to account for the heat load increase during the cavity voltage ramp-up without destabilizing the cavity. As a compensatory measure, the outlet cryogenic valve plug has been replaced in order to increase its inner flow coefficient (CV) to a value close to its maximum design capability\footnote{CV has been increased from 1.5 to 3.2 after plug replacement.}. This provided more margin for operation, particularly in the case of an increase in the heat loads. The operation of the cavity A11 during the year 2021 proved satisfactory, with stable behavior and no increase or decrease in heat load. However, the gain margin remained small. A significant degradation of the heat load would require additional measures, such as low accelerating field operation\footnote{This could mean to move the cryomodule to a different position in the beginning of the LINAC where the required accelerating gradients for the target current and energies are lower.} or replacement with a drift tube to allow repair.

\subsection{Next steps : Making operation smarter}
The cryogenic process at SPIRAL2 relies on several PLC-based control systems \cite{trudel:icalepcs2021-tupv006}. These systems drive the operation as well as other related systems, such as isolation vacuum and mechanical frequency tuning of the cavities. The previously described Linear Quadratic regulators were pre-tuned at a given set point in a separate MATLAB framework.  Their parameters are then embedded in the command-control PLCs. Day-to-day operations may require re-tuning of the LQ regulator parameters by operators. For this purpose, a new human–machine interface is being developed. This interface allows regulation specialists to 
\begin{itemize}
	\item re-compute the linearization of the cryomodule models around a new set-point, 
	\item tune the heat load observers and LQ controllers parameters (weights, filters, dynamics),
	\item simulate the performances of the regulators, 
	\item switch between several tuning presets, 
	\item transfer the model and regulation matrix to the cryomodules PLCs. 
\end{itemize}
The linearized models of the cryomodules had a certain bandwidth around the calculated setpoint. Observers can monitor heat load deviations from a given set point. The next step would be to preload several linearized models around different heat loads in the PLCs and manually switch pre-sets upon an important deviation detection. As the estimated heat load data produced by the observers grow more reliably, it might be possible to exploit them to provide more explicit information, such as warnings, alarms, or even to automatize the re-linearization process of the cryogenic model used to compute the observers and LQ controllers.
A further step, under development, would be to learn the nonlinear cryomodule models thanks to machine learning techniques using the twin cryogenic model and the operation data for increased precision. The models would then be embedded in their lightweight versions \footnote{see Deepgreen and LightML} in the control system.

\section{Conclusion}
SPIRAL2 has proven the value of advanced cryogenic and superconducting technologies to ensure the reliable operation of superconducting LINAC. The transition from commissioning to full operation has highlighted innovative solutions to complex challenges such as overcoming thermo-acoustic oscillations and using machine learning for cavity diagnostics. By incorporating thermodynamic models and optimized control systems, the facility achieves both stability and efficiency, even under demanding conditions.

As the focus shifts toward smarter and more adaptive systems, advanced techniques such as drift surveillance and automatic regulator calibration are planned for integration into command and control systems. However, a superconducting linear accelerator comprises interconnected and interdependent subsystems, including radiofrequency (RF), beam diagnostics, and cryogenics. This interdependency presents opportunities for smarter and more reliable operation through the use of generalized state observers and advanced diagnostic tools.
\bibliography{ref_prab}

\end{document}